\pdfoutput=1
\documentclass[11pt]{article}

\usepackage{amsmath, amsthm, amssymb}
\usepackage{hanging}

\begin{document}
\vspace*{0.25in}

\begin{center}
   {\Huge Is Arrow's Dictator a Drinker?}\\
   {\vspace*{8pt}\Large\em A Note on Misinterpretations of Arrow's Theorem}\\
   {\vspace*{16pt}\large Jeffrey Uhlmann}\\
   {\small Dept.\ of Electrical Engineering and Computer Science}\\
   {\small University of Missouri - Columbia}
\end{center}

\section{Introduction}

Kenneth Arrow's celebrated ``impossibility'' theorem establishes a set of conditions that cannot simultaneously hold for any ranked system of voting involving more than two alternatives \cite{arrow}. The significance of Arrow's theorem is that his conditions have strong intuitive appeal as fundamental requirements for any {\em reasonable} definition of a fair voting system. Arrow's result that they cannot hold simultaneously is so counterintuitive that it is sometimes referred to as Arrow's {\em paradox} because of its implication that there can be no ``fair'' voting system. This surprising negative conclusion has attracted significant attention not only in the field of social choice theory, but also more broadly as an example of how strongly-held intuitive expectations can be subverted when subjected to rigorous analysis.

In this note, we argue that the conventional interpretation of Arrow's theorem is fundamentally mistaken. This is not to say that it is incorrect -- {\em we find it to be entirely sound} -- rather we suggest that it does not imply an inconsistency with respect to the generally held intuitive notions of electoral fairness to which it purports to apply. 

\section{Arrow's Fairness Conditions}

For any rank-order voting system, Arrow's theorem says that the following three conditions cannot hold simultaneously:
\begin{enumerate}
\item If every voter prefers alternative $X$ over alternative $Y$, then the system ensures that its detemined electoral preference, i.e., the result of voting by the group, will be $X$ over $Y$.

\item If every voter's preference with respect to $X$ and $Y$ remains unchanged, then the electoral preference with respect to $X$ and $Y$ will be invariant to changes of preference with respect to other alternatives, e.g., if the electoral preference is $X$ over $Y$, and no voter has a change of preference with respect to $X$ and $Y$, then no other changes of preference with respect to either $X$ or $Y$ and some other alternative $Z$ will change the original electoral preference of $X$ over $Y$. 

\item There is no ``dictator'' whose vote determines the electoral preference.
\end{enumerate}

The first and third of these conditions are typically taken as self-evident requirements for any fair voting system, and a deeper reflection on the second condition typically reveals it to have commensurately strong intuitive appeal. The ``naturalness'' of these conditions may be appreciated even more by considering the implications if a voting system were to permit violations of any one of them. The surprising conclusion of Arrow's theorem is that one of these three conditions must be sacrificed because no system can simultaneously enforce all three. 

The three conditions given above are of course stated somewhat colloquially, but each can be formalized in a way that is amenable to analysis. For example, in the case of condition 3, the no-dictator condition, Arrow's definition can be stated as follows:
\begin{quote}
Let $V$ be the set of voters, $A$ the set of alternatives, and $f: V \times A \times A \rightarrow A$ the voting system's preference function, where $f(v,x,y) = x$ if and only if voter $v$ prefers $x$ over $y$. A voter $v$ is defined to be a dictator if for every pair of alternatives $x, y \in A$, the function $f$ holds.
\end{quote}

\section{The Drinker Paradox}

The drinker paradox relates to a scenario involving a non-empty pub, and a claim that there always exists a person such that if that person is drinking, then everyone is drinking \cite{drinker}. This can be stated more formally as follows:
\begin{quote}
Let $P$ be the set of pub patrons with function $g: P \times \{0, 1\} \rightarrow \{0, 1\} \) representing drinking status, where
$g(d, 1) = 1$ if and only if person $d$ is drinking. If there exists a person $d\in P$ for whom this function holds for everyone in $P$, then everyone in the pub is drinking.
\end{quote}
Technically, such a person always exists because whenever everyone in the pub is drinking, we can select any person in the pub and note that they are drinking -- {\em and that everyone else is  drinking as well}. The resolution of the paradox becomes immediately apparent when equivalently restated as follows: 
\begin{quote}
Whenever everyone in the pub is drinking, there exists a specific person who can be identified as drinking. 
\end{quote}
The counter-intuitiveness of the original formulation comes from the fact that it seems to suggest a causal relationship between the drinking of a particular person and the drinking of everyone else in the pub. We present the drinker paradox as an analogy to motivate our key observation:\\
~\\
\noindent {\bf Observation}: {\em Arrow's formal definition of a dictator does not imply the existence of a voter with power to influence the choices of other voters, so it does not comport in any way with the conventional definition of a dictator}.\\
~\\
This observation is important because if there is no element of coercion in Arrow's formal definition of a dictator, then his definition does not correspond to the intuitive notion of a dictator, and this severely undermines its intuitive appeal as a condition that should be expected of any fair voting system. In other words, use of the word ``dictator'' in the informal description of Arrow's third condition connotes the existence of a voter with some kind of coercive power over other voters, which is clearly inconsistent with a fair voting system, but no such coercive influence is captured by the formal definition applied by Arrow. It can now be seen that Arrow's dictator is as much an artifact of formulation as the seeming influence of a special pub-goer in the drinker paradox. As was seen with the drinker paradox, we can identify an equivalent formulation of Arrow's theorem that is more intuitively revealing:
\begin{quote}
Given the first two fairness conditions, it is impossible to prevent the possible occurrence of a voter $v$ for whom $f$ holds for every pair of alternatives $x, y \in A$.
\end{quote}
Now it is more clearly seen that $v$ can be pivotal with respect to $f$ purely ``by chance,'' so it should not be surprising that no guarantee can be made that such a voter can never occur.   

\section{Reconsidering the Scope of Arrow's Theorem}

A natural response to the observation of this paper would be to suggest that it is merely an issue of semantics and does not in any way undermine the correctness or significance of Arrow's theorem -- which is a cornerstone of social choice theory that has been examined extensively in a wide range of applications for well over a half a century. However, such a position is untenable for the following reasons:
\begin{description}
\item[Correctness]- While it is true that Arrow's theorem follows from its definitions, and thus is technically correct in that sense, if it is based on an incorrect formulation of the problem, then conclusions derived from that formulation cannot be correctly applied to that problem. 

\item[Significance]- If the no-dictator condition is in fact formalized in a way that is overly inclusive, i.e., encompasses scenarios that do not involve a voter with coercive power over other voters, then its incompatibility with the other two conditions is of purely academic interest with no clear relevance to conventional notions of fairness. 
\end{description}
More concretely, let $S$ be a rank-order voting system that satisfies Arrow's first two conditions and formally assumes that voters are not subject to coercion, i.e., they are assumed to be unrestricted in their ability to express their true preferences when voting, then $S$ can be fair according to the standard of Arrow. This is because the absence of coercion (i.e., no {\em true} dictator) is an assumption that is independent of the first two conditions, which Arrow accepts are not mutually inconsistent.

\section{Discussion}

In this note we have argued that Arrow's theorem is founded on a misformulation of the no-dictator condition. Specifically, while Arrow's formal definition of a dictator encompasses scenarios that are {\em consistent} with the existence of coercion by a dictator, it is not {\em exclusive} to such scenarios. We have intuitively illustrated this issue by analogy to the drinker paradox, which defines a scenario that is suggestive of a causal influence of one person on the behavior of others when in fact the behaviors are simply correlated due to structural properties of the formulation of the scenario.

Two positive conclusions can be drawn from this exercise. The first is that the possibility of a fair rank-voting system is not precluded by Arrow's theorem. The second is encouragement that it may be fruitful to carefully examine the scope of applicability of other prominent counterintuitive theorems -- {\em even ones that have enjoyed longstanding universal acceptance}. More generally, we suggest that the assumptions underpinning a given result should receive scrutiny that is at least inversely proportional to the intuitiveness of the result.

\bibliographystyle{amsplain}

\end{document}